\documentclass[12pt]{elsart}

\usepackage{graphicx,epsfig,amsmath,amssymb,amsfonts}
\usepackage{latexsym,cite,eucal}
\newcommand{\beq}{\begin{equation}}
\newcommand{\eeq}{\end{equation}}
\newcommand{\nn}{\nonumber \\}
\newcommand{\fs}{\,.}
\newcommand{\co}{\,,}

\renewcommand{\dag}{^\dagger}

\newcommand{\eps}{\epsilon}

\newcommand{\MeV}{\,\mbox{MeV}}

\newcommand{\mpn}{M_{\pi^0}}
\newcommand{\mpc}{M_\pi}

\newcommand{\Fnr}{\mathcal{F}}
\newcommand{\mpr}{m_p}
\newcommand{\mn}{m_n}

\newcommand{\vq}{{\bf q}}
\newcommand{\vk}{{\bf k}}
\newcommand{\kt}{\alpha_{0,0}}
\newcommand{\knlq}{\alpha_{1,0}}
\newcommand{\knlk}{\alpha_{0,1}}
\newcommand{\qstar}[1]{h^{#1}_{cd}}

\newcommand{\ratio}{\eta}

\begin{document}

\begin{frontmatter}

\title{\Large\bf Pion electroproduction in a nonrelativistic theory}

\author{Andreas Fuhrer$^1$}

\address{University of California, San Diego
\\9500 Gilman Drive, La Jolla, CA 92093--0319}

\thanks[fuhrer]{afuhrer@physics.ucsd.edu}

\begin{abstract}
A nonrelativistic effective theory to describe the electroproduction
reaction of a single pion on the nucleon at leading order in the
electromagnetic coupling is constructed. 
The framework is tailored to accurately describe the cusp generated by the pion
and nucleon mass differences. 
The $S$- and $P$- wave multipole amplitudes at two loops for all four reaction
channels are provided.
As an application, a new low energy theorem is discussed.
\end{abstract}

\begin{keyword}
 Chiral symmetries \sep Meson production \sep Pion-baryon interactions 
\PACS 11.30.Rd, 13.60.Le, 13.75.Gx 
\end{keyword}
\end{frontmatter}

{\bf 1.} Pion electroproduction is one of the reactions in particle physics
which has been studied ever since almost the beginning of the
field. Since the first experiment over fifty years ago \cite{ex0}, the
accuracy has been constantly increased and the threshold region at low
values of the photon virtuality became accessible
\cite{ex1,ex2,weis}. It has been known for a long time that at low
energies, this process allows one to study pion-nucleon physics. 
The strength of the pronounced cusp in the
production channel of neutral pions
which appears at the $n \pi^+$ threshold is intimately related to the
pion-nucleon scattering lengths \cite{FW}. 
One of the aims of the present
letter is to derive this connection in an effective field theory with
a suitable perturbative expansion. The standard tool to study any hadronic
low-energy process in the Standard Model is chiral perturbation theory (ChPT)
\cite{Weinberg,GL1983,GL1985,GSS}, which has been applied to the
reaction in question in a series of articles
\cite{BKMep1,BKMep2,BKMep3,BKMep4}. Here, we propose a nonrelativistic effective theory
which has already been successfully applied to hadronic atoms (see
Ref.~\cite{ha} and references therein), $K,\eta,\eta' \to 3\pi$
decays \cite{CGKR,BFGKR,KS}, $K_{e4}$ decays \cite{CGR} and
which was recently also formulated for pion photoproduction
\cite{af}. 

The strength of the nonrelativistic framework in comparison to ChPT
is that the fundamental quantities of pion-nucleon interactions at low
energies, i.e. the coefficients of the effective range expansion
(scattering lengths, effective range parameters, ...), are {\it free parameters}
of the theory. There is no expansion of these parameters in terms of
$\mpc/\Lambda_{QCD}$ or $\mpc/\mpr$. On the other hand, the drawback is a
more restricted region of validity than in ChPT.
However, it was noted in Ref.~\cite{BKMep2} that especially for the
$S$-wave multipoles, the expansion in terms of $\mpc/\mpr$ converges
rather slowly. Therefore, a limited region of applicability might be a
price worth to pay to avoid this expansion.\newline
The letter is organized as follows: After collecting the notation, the
kinetic part of the Lagrangian is written down and the power counting
is discussed. Then the interaction terms are constructed and the
matching relations of the coupling constants are discussed. The calculation of the multipoles up to and
including two-loop corrections is then straightforward. The letter
concludes with a discussion of a low energy theorem.

{\bf 2.} We calculate the electroproduction reaction at leading order in the
electromagnetic coupling. Therefore, only one photon is exchanged
between the electron and the nucleon and all the hadronic physics that
we are interested in is contained in the transition matrix element for
the process  $N(p_1)+\gamma^\star(k) \to N(p_2) + \pi^{0,\pm}(q)$,
where $\gamma^\star$ denotes an off-shell photon with $k^2 < 0$. A
discussion of pion electroproduction including the leptonic part can
be found for instance in Ref.~\cite{aff}.
In the following, the four reaction
channels will be abbreviated as
\begin{align}
\gamma^\star p \to p \pi^0 &: (p0) \co &\gamma^\star p \to n \pi^+ &: (n+) \co
&\gamma^\star n \to n \pi^0 &: (n0) \co &\gamma^\star n \to p \pi^-  &: (p-)\fs \nonumber
\end{align}
All the observables of electroproduction experiments can be expressed in terms of
electric, magnetic and scalar multipoles. The transition current
matrix element $\epsilon_\mu J^\mu$, where $\epsilon_\mu$ is the
photon polarization vector, is conveniently written in terms of two component spinors $\xi_t$ and Pauli matrices
$\tau^k$ \cite{bdw} and Coulomb gauge, $\nabla {\bf A} = 0$, is
chosen,\footnote{The $\Fnr_{5,6}$ defined like that are sometimes also
called $\Fnr_{7,8}$ in the literature.}
\begin{align}\label{eq:nrr}
\epsilon_\mu J^\mu &= \xi\dag_{t'}\, \Fnr\,
\xi_{t} \co \nn 
\Fnr &= i \boldsymbol\tau \cdot \boldsymbol\epsilon\,  \Fnr_1+
\boldsymbol\tau\cdot \hat{{\bf q}}\, \boldsymbol\tau\cdot (\hat{{\bf k}}\times
\boldsymbol\epsilon)\,\Fnr_2 +i \boldsymbol\tau\cdot \hat{{\bf k}}\,
\hat{{\bf q}}\cdot \boldsymbol\epsilon\, \Fnr_3 +i \boldsymbol\tau
\cdot\hat{{\bf q}}\, \hat{{\bf q}}\cdot \boldsymbol\epsilon\,
\Fnr_4\nn
& -
i \boldsymbol\tau \cdot \hat{{\bf k}}\, \epsilon_0\, \Fnr_5 - i
\boldsymbol\tau \cdot \hat{{\bf q}}\, \epsilon_0\, \Fnr_6
\fs
\end{align}
The hat denotes unit vectors. The $\Fnr_i$ are decomposed into
electric, magnetic and scalar multipoles with the help of 
derivatives of the Legendre polynomials $P_l(z)$ \cite{bdw},
\begin{align}
\Fnr_1 &= \sum_{l=0}\,
    [lM_{l+}+E_{l+}]P_{l+1}'(z)+[(l+1)M_{l-}+E_{l-}]P_{l-1}'(z) \co \nn
\Fnr_2 &= \sum_{l=1}\,[(l+1)M_{l+}+lM_{l-}]P_l'(z) \co \nonumber
\end{align}
\begin{align}
\Fnr_3 &=
\sum_{l=1}\,[E_{l+}-M_{l+}]P_{l+1}''(z)+[E_{l-}+M_{l-}]P_{l-1}''(z)
\co \nn
\Fnr_4 &= \sum_{l=1}\,[M_{l+}-E_{l+}-M_{l-}-E_{l-}]P_l''(z) \co \nn
\Fnr_5 &= \sum_{l=0}\,[l S_{l-}-(l+1)S_{l+}] P_l'(z) \co \nn
\Fnr_6 &= \sum_{l=1}\,[(l+1)S_{l+}P_{l+1}'(z) - l S_{l-} P_{l-1}'(z)] \fs
\end{align}
The multipoles are complex valued functions of the center of
mass energy and the photon virtuality.
The discussion is restrained to the
center of mass frame in the rest of the article.

{\bf 3.} We aim at a description of the multipoles
close to the reaction threshold --- where the momentum of the produced
pion and of the nucleon is small. A nonrelativistic theory is the
right tool for this task. The other kinematic variable, the
photon virtuality $k^2$, has to be restricted to small values compared
to $4\mpc^2$. This allows for an expansion of the amplitudes
in $k^2/(4\mpc^2)$. The factor of four shows up because a
power series in $k^2$ around the origin converges inside a circle in the complex plane
up to the first branch point which lies at $k^2 =
4\mpc^2$. A nonrelativistic treatment
also offers the advantage that
all the masses can be set to their physical value. Therefore, all the
poles and branch points appear at the {\it correct place} in the Mandelstam
plane. Moreover, the interaction of the nucleon and the pion is described by effective
range parameters, which allows one to directly access the pion-nucleon scattering
lengths, the main goal of the present analysis. 

The covariant formulation of nonrelativistic field theories introduced
and applied in Refs.~\cite{CGKR,BFGKR,KS,af} is used here. It
incorporates the correct relativistic dispersion law for the
particles. Note however that it is not mandatory to keep all the
higher order terms, but merely a matter of convenience. The
nonrelativistic proton, neutron and pion fields are denoted by $\psi$,
$\chi$ and $\pi_k$, respectively. The kinetic part of the Lagrangian
after minimal substitution takes 
the form (see Ref.~\cite{BFGKR2})
\begin{align}
\CMcal{L}_{kin} &= \sum_\pm\Bigl(i\pi_\pm^\dagger D_t{\cal W}_\pm\pi_\pm
-i(D_t{\cal W}_\pm\pi_\pm)^\dagger\pi_\pm-2\pi_\pm^\dagger {\cal W}_\pm^2\pi_\pm\Bigr)+ i\psi^\dagger D_t{\cal W}_p\psi\\
&-i(D_t{\cal W}_p\psi)^\dagger\psi-2\psi^\dagger {\cal W}_p^2\psi + 2
\chi\dag W_n (i \partial_t-W_n)\chi + 2 \pi_0\dag W_0
(i\partial_t-W_0)\pi_0 \co \nonumber
\end{align}
with
\begin{align}
W_0 &= \sqrt{\mpn^2-\triangle} \co &W_n &= \sqrt{m_n^2-\triangle}\co
&D_t\pi_\pm &= (\partial_t\mp ieA_0)\pi_\pm \co\nn
D_t\psi &= (\partial_t- ieA_0)\psi \co &{\cal W}_\pm &= \sqrt{M_\pi^2-{\bf D}^2} \co &{\cal W}_p &=
\sqrt{\mpr^2-{\bf D}^2} \co \nn 
{\bf D}\pi_\pm &= (\nabla\pm ie{\bf A})\pi_\pm \co &
{\bf D}\psi &= (\nabla + ie{\bf A})\psi \fs
\end{align}
Since the photon is treated as an external field, its
kinetic term is absent.

{\bf 4.} Close to threshold, the momenta of the outgoing pion and
the outgoing proton, normalized with the pion mass, are small and
therefore counted as  a quantity of
$O(\eps)$. The normalized momenta of the incoming proton and of the
photon are counted as
$O(1)$. All the masses are counted as $O(1)$. The mass differences
of the charged and neutral pion, $\Delta_\pi/\mpc^2 \equiv (\mpc^2-\mpn^2)/\mpc^2$ and of
the proton and the neutron, $\Delta_N/\mpc^2 \equiv (\mn^2-\mpr^2)/\mpc^2$ are counted as $O(\eps^2)$.
The power counting therefore is exactly the same as in the case of
photoproduction, Ref.~\cite{af}. The
only difference is that the off-shell photon introduces an additional scale into
the problem, the virtuality $k^2$, which is taken to be small in comparison
to $4\mpc^2$ by assumption. Therefore, the quantity $k^2/(4\mpc^2)$ is
also counted as $O(\epsilon^2)$. 
This power counting is valid although
derivatives on the incoming nucleon and photon fields generate terms
of $O(1)$. Simply expand the large momenta in $\epsilon$,
\begin{align} \label{eq:k}
|{\bf k}| &= \sum_{n,m}\alpha_{n,m} \vq^{2n} k^{2m} \co &\kt &=
\frac{\mpn}{2}\, \frac{2+y}{1+y} \co &\knlq &= \frac{y^2+2y+2}{4 \mpn
  (1+y)} \co \nn
\knlk &= -\frac{y^2+2y+2}{2\mpn(1+y)(2+y)} \co &y &=
\frac{\mpn}{\mpr} \co
\end{align}
and define the coupling of the leading order term such that it
contains all the large terms with $\alpha_{0,0}$ (for more details, see Ref.~\cite{af}).
The derivatives on the incoming fields are only
needed to generate unit vectors in the direction of the incoming
photon.\newline
The rescattering of the pion on the nucleon introduces another expansion
parameter $|\vq|^n b$, where $b$ denotes an arbitrary effective range parameter
 of $\pi N$ scattering (at higher orders, $b$ can also contain masses,
 see for instance Eq.~(\ref{eq:matchingD}))
 with $n$ the pertinent number to obtain a dimensionless
 quantity. In the following, I will indicate the
 order of a given quantity $W$ as $a^m \epsilon^n$, where $m$ simply counts the
 number of loops and $n$ denotes the combined power of momenta $\vq$
 and $k^2$ present in $W$. It
 is straightforward to see how each of the terms divides up into the
 different dimensionless expansion parameters $|\vq|^nb$  and $\epsilon$. 
It was found in Ref.~\cite{af} that the expansion in $|\vq|/\mpc$
works well at
least up to $|\vq| \simeq 70 \MeV$. The expansion in $|\vq|^n b$ is
expected to converge even faster, since the effective range parameters
are much  smaller than $1$ in units of inverse $\mpc$.

{\bf 5.} The Lagrangian needed for the calculation of the amplitudes for pion 
electroproduction reads $\CMcal{L} = \CMcal{L}_{kin}+\CMcal{L}_\gamma+\CMcal{L}_{\pi N}$,
where $\CMcal{L}_{kin}$ denotes the kinetic part, $\CMcal{L}_\gamma$
incorporates the interaction with the photon field and 
$\CMcal{L}_{\pi N}$ describes the
pion-nucleon sector.\newline
For the pion nucleon sector, the Lagrangian was given before in
Refs.~\cite{LR,af}. For convenience, we write it down again. An
equivalent description can be found in Ref.~\cite{bs} (see
Ref.~\cite{af} for an explicit comparison).
For every channel $n$, we collect the charges of the outgoing and the
incoming pions in the variables $v$ and $w$, $(n;v,w)$: $(0;0,0) ,
(1;0,+),\, (2;+,+),\, (3;0,0),\, (4;-,0),\, (5;-,-)$, thereby assigning 
unique values to the variables $v$ and $w$ once $n$ is given. The
Lagrangian reads
\begin{align}\label{eq:Lpn}
\CMcal{L}_{\pi N} &= \left( \psi\dag \,\, \chi\dag \right)
\left( \begin{array}{cc} T_{\{0,5 \}} & T_{\{1,4\}}
  \\ T_{\{1,4\}}\dag & T_{\{2,3 \}} \end{array}
\right) \left( \begin{array}{c} \psi \\ \chi \end{array} \right) \co
\\[1mm]
T_\CMcal{C} &= \sum_{n \in\, \CMcal{C}}\left[ C_n \pi_v\dag \pi_w + D_n^{(1)}\nabla^k
\pi\dag_v \nabla^k \pi_w + D_n^{(2)} \pi\dag_v
\overleftrightarrow{\triangle} \pi_w + i D_n^{(3)} \tau^k
\epsilon^{ijk}\nabla^i \pi\dag_v \nabla^j \pi_w \right]  \nonumber
\end{align}
with the abbreviation $f\overleftrightarrow{\triangle}g \equiv f \triangle
g + (\triangle f) g$.\newline
For $\CMcal{L}_\gamma$, the photon is treated as an external
field. 
Gauge invariance requires that $A^\mu$ can 
only appear in covariant derivatives and through the Maxwell equations in
the electric and magnetic fields ${\bf E} =
-\boldsymbol\nabla A^0 - \dot{{\bf A}}$ and ${\bf B} =
\boldsymbol\nabla \times {\bf A}$. 
All the terms are invariant under space rotations,
parity and time reversal transformations.
The upper index on the coupling constants counts
the number of derivatives on the external vector field and is
introduced for later convenience. The terms with coupling constants
$G_i^{(k)}$ with $i = 0,\ldots,15$ were already given in Ref.~\cite{af} and
contribute to the photoproduction amplitudes.
\begin{align}
\CMcal{L}^{(0)}_\gamma &= -i G_0^{(1)}\psi\dag \tau^k \psi\, E^k\,
\pi_0\dag \nn
\CMcal{L}^{(1)}_\gamma &= -i G_1^{(2)}\, \psi\dag \tau^k
\psi\, \nabla^j E^k\, \nabla^j \pi_0\dag  + i G_2^{(1)}\, \psi\dag \tau^m \tau^l
\psi\,B^l\, \nabla^m \pi_0\dag \nn
 &-i G_3^{(2)}\, \psi\dag \tau^j \psi\,
\nabla^j E^k\, \nabla^k \pi_0\dag -i G_{20}^{(2)}\psi\dag \tau^k \psi
\nabla^j E^j \nabla^k\pi_0\dag  \nn
\CMcal{L}^{(2)}_\gamma &= -iG_4^{(3)} \psi\dag \tau^k \psi \nabla^{jl} E^k
\nabla^{jl} \pi_0\dag -i G_5^{(1)} \psi\dag \tau^k \psi E^k \triangle \pi_0\dag \nn
&+i G_6^{(2)} \psi\dag \tau^m \tau^l \psi \nabla^{n} B^l \nabla^{mn}
\pi_0\dag  -i G_7^{(3)} \psi\dag \tau^j \psi \nabla^{jl} E^k \nabla^{kl}
\pi_0\dag  \nn 
&-i G_8^{(1)} \psi\dag \tau^j \psi E^k \nabla^{jk}\pi_0\dag
-iG_{16}^{(3)} \psi\dag \tau^k \psi \CMcal{P}E^k \pi_0\dag \nn
&-i G_{21}^{(3)} \psi\dag \tau^k \psi \nabla^{mj}E^j \nabla^{mk}
\pi_0\dag \nn
\CMcal{L}^{(3)}_\gamma &= -iG_9^{(2)} \psi\dag \tau^k \psi \nabla^j E^k
\triangle \nabla^j \pi_0\dag -i G_{10}^{(4)} \psi\dag \tau^k \psi
\nabla^{lmn}E^k \nabla^{lmn} \pi_0\dag \nn &+ i G_{11}^{(1)} \psi\dag \tau^m
\tau^l \psi B^l \triangle \nabla^m \pi_0\dag + i G_{12}^{(3)}
\psi\dag \tau^m \tau^l \psi \nabla^{in} B^l \nabla^{min} \pi_0\dag
\nn
&-i G_{13}^{(2)} \psi\dag \tau^j \psi \nabla^j E^k \triangle \nabla^k
\pi_0\dag -i G_{14}^{(4)} \psi\dag \tau^j \psi \nabla^{jlm} E^k \nabla^{klm}
\pi_0\dag \nn
&-i G_{15}^{(2)} \psi\dag \tau^j \psi \nabla^l E^k \nabla^{jkl}
\pi_0\dag -i G_{17}^{(4)} \psi\dag \tau^k \psi \nabla^j\CMcal{P} E^k
\nabla^j \pi_0\dag\nn
&- i G_{18}^{(4)} \psi\dag \tau^j \psi \nabla^j \CMcal{P} E^k
\nabla^k \pi_0\dag+iG_{19}^{(3)} \psi\dag \tau^m \tau^l \psi \CMcal{P}
B^l \nabla^m \pi_0\dag \nn
&-i G_{22}^{(2)} \psi\dag \tau^k \psi \nabla^j E^j \triangle \nabla^k
\pi_0\dag-iG_{23}^{(4)} \psi\dag \tau^k \psi \nabla^j\CMcal{P} E^j
\nabla^k \pi_0\dag
\end{align}
Here, the notation $\nabla^{i_1i_2\ldots i_k} \equiv \nabla^{i_1}
\nabla^{i_2} \cdots \nabla^{i_k}$ and $\CMcal{P} \equiv -\partial_t^2+\triangle$ is
introduced. Since the structure of the
Lagrangian for the remaining channels stays the same, one only has to
replace the coupling constants and the field operators,
\begin{align}\label{eq:rep}
(n+) &: \{\psi\dag,\pi_0\dag,G_i^{(n)}\} \to \{\chi\dag,\pi_+\dag,H_i^{(n)} \} \co
  &(n0) &: \{\psi,\psi\dag, G_i^{(n)} \} \to \{ \chi,\chi\dag,L_i^{(n)}\}  \co \nn
  (p-) &: \{\psi,\pi_0\dag,G_i^{(n)} \}   \to \{\chi,\pi_-\dag,  K_i^{(n)}\}
  \fs
\end{align}
The full Lagrangian $\CMcal{L}_\gamma$ is given by
adding the $\CMcal{L}^{(i)}_\gamma$ of all four channels.

{\bf 6.} In the pion-nucleon sector, the coupling constants of the
nonrelativistic Lagrangian, $C_i$ and $D^{(k)}_i$ can be expressed in
terms of pion-nucleon scattering lengths of the $S$-wave and $P$-wave,
$a_{0+}$ and $a_{1\pm}$ and the effective range parameter $b_{0+}$, respectively. Adopting the notation of
Ref.~\cite{Hohler},
in the isospin limit, the isospin decomposition of the $\pi N$ scattering amplitudes reads
\begin{align}\label{eq:amps}
T_{p\pi^0\to p\pi^0} &= T_{n\pi^0 \to n\pi^0} = T^+ \co &T_{p \pi^0
  \to n\pi^+} &= T_{n\pi^0 \to p
  \pi^-} = -\sqrt{2}\, T^-
\co \nn T_{n\pi^+ \to n \pi^+} &= T_{p\pi^- \to p\pi^-} = T^++T^- \fs
\end{align}
Defining $\CMcal{N} = 4\pi (\mpr+\mpc)$, one finds
\begin{align}\label{eq:matching}
C_0 &= 2\, \CMcal{N}  a_{0+}^+ \co &C_1 &= 2\sqrt{2}\,\CMcal{N}
a_{0+}^-  \co &C_2 &= 2\, \CMcal{N} (a_{0+}^++a_{0+}^-) \co\nn
C_3 &= C_0 \co &C_4 &= C_1 \co &C_5 &= C_2 \fs
\end{align} 
The matching conditions for the $D_i^{(k)}$ are given in a generic
form only. The isospin index of the threshold parameters can be
inferred from Eq.~(\ref{eq:amps})\footnote{The Condon-Shortley phase
convention is used.}.
\begin{align}\label{eq:matchingD}
D^{(1)}_i &= 2\CMcal{N} (2 a_{1+}+a_{1-}) \co &D^{(2)}_i &= -
\CMcal{N} \left(\frac{a_{0+}}{2\mpr \mpc}+b_{0+} \right) \co \nn
D^{(3)}_i &= 2\CMcal{N} ( a_{1-}-a_{1+})\fs
\end{align}
Here, higher order terms in the threshold parameters have been dropped.
The corrections to these relations which appear due to isospin breaking have to be
calculated within the underlying relativistic theory. For the $C_i$, they can be found
in Refs.~\cite{GILMR,MRR,hkm}. In the isospin limit, the different
couplings $C_i$ are related according to
Eq.~(\ref{eq:matching}). These relations do not hold anymore once
isospin breaking corrections are taken into account.

The multipole coupling constants $G^{(n)}_i$, $H^{(n)}_i$, $K^{(n)}_i$ and $L^{(n)}_i$ on the other hand are related to the
threshold parameters of the electric and magnetic multipoles of the
pertinent channel. 
In the isospin limit, the expansion of the real part of the multipole
$X_{l\pm}$ close to threshold is written in the form 
\begin{align}\label{eq:thresholdpar}
\mathrm{Re}X_{l\pm}(s,k^2) &= \sum_{k,m} \bar{X}_{l\pm,2k,2m}|\vq|^{l+2k}k^{2m} \co
\end{align}
which defines the threshold parameters $\bar{X}_{l\pm,2k,2m}$. In the
following, the relations of the coupling constants $G^{(n)}_i$ to these threshold
parameters is given dropping terms of the order of the pion-nucleon threshold
parameters. Since the nonrelativistic theory is not suited for the study of the dependence
of the multipoles on $|\vk|$, in this analysis, all vectors $\vk$ are turned into unit
vectors by the pertinent redefinition of the coupling constants,
\beq\label{eq:redef}
 G_i^{(n)} = G_i/\kt^n \fs 
\eeq
The higher order corrections due to factoring out $|\vk|$ are taken care
of in the matching relations.
Again, these relations pick up isospin breaking corrections which have to be
evaluated in the underlying relativistic theory.

Only the matching equations for the couplings of the Lagrangians
$\CMcal{L}_\gamma^{(0)}$ and $\CMcal{L}_\gamma^{(1)}$ are indicated
in the main text, the remaining relations are relegated to appendix
\ref{app:matching}. To ease notation, $\bar{X}_{i\pm} \equiv
\bar{X}_{i\pm,0,0}$ is used.\footnote{The matching differs from the
  photoproduction case given in Ref.~\cite{af} because the amplitudes
  are normalized differently.}
\begin{align}
G_0 &= \bar{E}_{0+} \co  &G_1 &= 3(\bar{E}_{+1}+\bar{M}_{+1})
\co \nn
G_2 &= -2\bar{M}_{1+}- \bar{M}_{1-} \co  &G_3 &= 3
(\bar{E}_{1+}-\bar{M}_{1+}) \fs
\end{align} 
For the coupling constants $H_i$, $K_i$ and $L_i$, the algebraic form of the relations is identical. However, the multipoles of
the pertinent channels appear and the masses in Eq.~(\ref{eq:redef})
have to be adjusted. In the isospin limit, the leading multipole
couplings fulfill
\begin{align}
\sqrt{2}\left( G_0 - L_0\right) &= H_0 + K_0 \fs
\end{align}
Since the effects of dynamical photons are not discussed here, all the coupling
constants are taken to be real\footnote{A more detailed discussion can
  be found in Refs.~\cite{af,hkm,hkm2}.}.

{\bf 7.} In the following, we provide the expressions for the electric,
magnetic and scalar multipoles $E_{l+}$, $S_{l\pm}$ and $M_{l \pm }$
for the channel $(p0)$\footnote{We refrain from using an additional index on
the multipoles to indicate the channel. Only when tree level
coefficients of a channel different from $(p0)$ appear in the result, this will be
indicated.}.
The result is written in the form 
\begin{align}\label{eq:result}
X_{l,\pm}(s) &= X_{l\pm}^{\mathrm{tree}}(s)+X_{l\pm}^{\mathrm{1
    Loop}}(s) +X_{l\pm}^{\mathrm{2
    Loop}}(s) \cdots
\end{align}
where $s = (p_1+k)^2$ and the ellipsis denote higher order terms in
the perturbative expansion.
The results for the other channels can be recovered by a simple replacement of the coupling
constants which will be given later. Write
\begin{align}\label{eq:tree}
X^{\mathrm{tree}}_{l\pm}(s) &= \vq^l \left[ X^t_{l\pm}+ X^t_{l\pm,q}
\vq^{2} + X^t_{l\pm,k} \frac{k^2}{\kt^2} +
X^t_{l\pm,\ratio} \eta + \cdots \right]
\end{align}
where  $\ratio = \frac{\knlk}{\alpha_{0,0}}k^2 +
\frac{\knlq}{\alpha_{0,0}} \vq^2$ collects the terms which are due to
the expansion of $k^0$ and ${\bf k}$ in $\epsilon$. One finds for the
leading terms
\begin{align}
E^t_{0+} &= G_0 \co & S^t_{0+} &= G_0 \co &6 E^t_{1+} &= G_1+G_3 \co\nn
3 M^t_{1-} &= G_3-G_1-3G_2 \co &6 M^t_{1+} &= G_1-G_3  \co &6 S^t_{1+}
&= G_1+G_3 \co\nn
3S^t_{1-} &= G_1+G_3+3G_{20}\co
\end{align}
for the coefficients of $\vq^2$ and $k^2$
\begin{align}
3 E^t_{0+,q} &= G_4-3 G_5+G_6-G_8 \co \nn
2 E^t_{0+,k} &= G_0+2 G_{16}\co \nn
3S^t_{0+,q} &= G_4-3 G_5+G_7-G_8+G_{21} \co \nn
S^t_{0+,k} &= G_{16} \co \nn
30E^t_{1+,q} &=
 -5G_9+3G_{10}+2G_{12}-5G_{13}
 +G_{14}-2G_{15} \co \nn
12 E^t_{1+,k} &= G_1+G_3+2G_{17}+2G_{18} \co \nn
30M^t_{1+,q} &= -5G_9+3G_{10}+2G_{12}+5G_{13}-G_{14} \co \nn
12 M^t_{1+,k} &= G_1-G_3 +2G_{17}-2G_{18} \co\nn
15 M^t_{1-,q} &=
5G_9-3G_{10}+15 G_{11}-5G_{12}-5G_{13}+G_{14} \co \nn
6 M^t_{1-,k} &= -G_1+G_3-2G_{17}+2G_{18} -6G_{19} \co \nn
30 S^t_{1+,q} &=
-5G_9+3G_{10}-5G_{13}+3G_{14}-2G_{15}  \co\nn
6  S^t_{1+,k} &= G_{17}+G_{18} \co\nn
15S^t_{1-,q} &= -5G_9+3G_{10}
-5G_{13}+3G_{14}-5G_{15}-15G_{22} \co\nn
3  S^t_{1-,k} &= G_{17}+G_{18}+3G_{23} \co 
\end{align}
and for the coefficients of $\eta$
\begin{align}
E^t_{0+,\ratio} &= G_0\co &S^t_{0+,\ratio} &= G_0\co &3
E^t_{1+,\ratio} &= G_1+G_3 \co \nn
3 M^t_{1-,\ratio} &= -2G_1-3G_2+2G_3 \co &3 M^t_{1+,\ratio} &= G_1-G_3 \co  &3S^t_{1+,\ratio}
&= G_1+G_3 \co \nn
3 S^t_{1-,\ratio} &= 2G_1 + 2G_3 + 6 G_{20} \fs
\end{align}
The coefficients $X^t_{l\pm}$ and $X^t_{l\pm,\ratio}$ are not equal
because the term with the coupling constant $G_2$ in the Lagrangian
does not have a time derivative on the photon field. 

{\bf 8.} All the one-loop contributions are proportional to the basic integral
\begin{align}
J_{ab}(P^2) &= \int \frac{d^Dl}{i(2\pi)^D} \frac{1}{2 \omega_a({\bf
    l})2\omega_b({\bf P}-{\bf l}) }\, \frac{1}{( \omega_a({\bf
    l})-l_0 ) (\omega_b({\bf P}-{\bf l}) -P_0 +l_0 ) } \co \nonumber
\end{align}
\begin{align}
\omega_\pm({\bf  p}) &= \sqrt{\mpc^2+ {\bf p}^2} \co  &\omega_i({\bf p}) &=
\sqrt{m_i^2+ {\bf p}^2}\co \qquad i=n,p  \nn
\omega_0({\bf p})  &= \sqrt{\mpn^2+ {\bf p}^2}\co  &P^2 &= P_0^2-{\bf
  P}^2 \fs
\end{align}
In the limit $D \to 4$,
\beq\label{eq:loopfunc}
J_{ab}(P^2) = \frac{i}{16\pi
  s}\sqrt{(s-(m_a+M_{\pi^b})^2)(s-(m_a-M_{\pi^b})^2)}\co 
\eeq
which is a quantity of order $\eps$.
\begin{figure}
\centering
\begin{tabular}{cc}
\includegraphics[height=1.4cm]{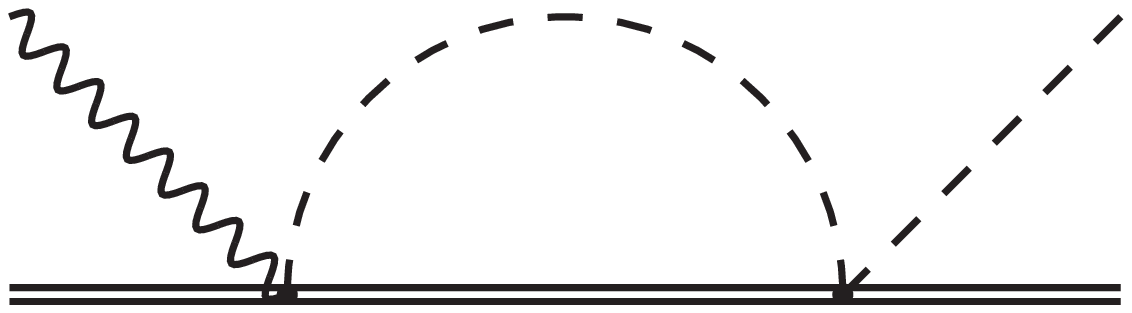}&\includegraphics[height=1.4cm]{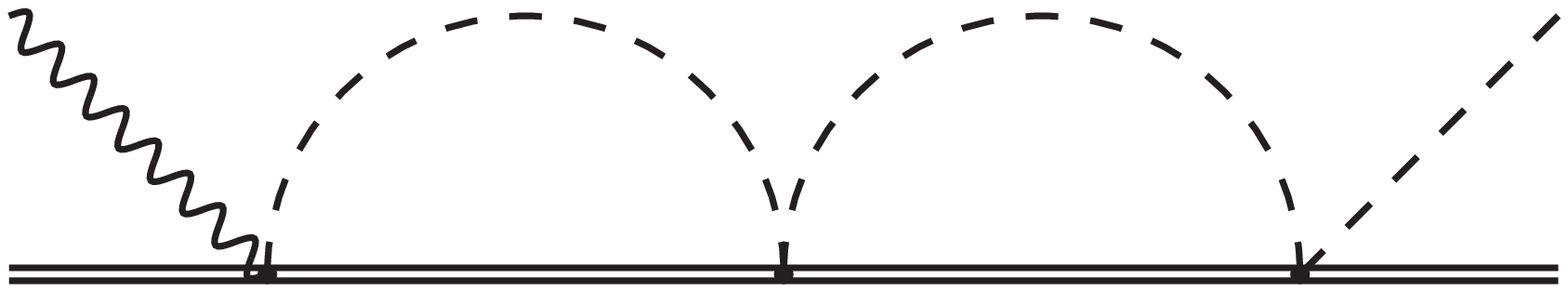}
\end{tabular}
\caption{One- and two loop topologies needed to calculate the
  amplitude. The double line
  generically denotes a nucleon, the dashed line a pion and the
  wiggly line indicates the external electromagnetic field.}\label{fig:diags}
\end{figure}
The one-loop result for channel $(c)$ up to and including order $O(a \epsilon^4)$ reads
\begin{align} \label{eq:1loop}
\left( \begin{array}{c} E_{0+}^{\mathrm{1 Loop}}(s) \\ \frac{1}{|\vq|}M_{1+}^{\mathrm{1
      Loop}}(s) \\ \frac{1}{|\vq|} M_{1-}^{\mathrm{1 Loop}}(s)
  \\ \frac{1}{|\vq|} E_{1+}^{\mathrm{1 Loop}}(s) \\ S_{0+}^{\mathrm{1
      Loop}}(s) \\ \frac{1}{|\vq|} S_{1+}^{\mathrm{1 Loop}}(s) \\ \frac{1}{|\vq|} S_{1-}^{\mathrm{1 Loop}}(s)  \end{array}
  \right) &= \left( \begin{array}{cc}  P^{(c)}_{11} & P^{(c)}_{12}\\ P_{21}^{(c)}  & P^{(c)}_{22} \\ P^{(c)}_{31} &
     P^{(c)}_{32} \\ P^{(c)}_{41} &
     P^{(c)}_{42} \\P^{(c)}_{51} & P^{(c)}_{52} \\ P^{(c)}_{61} &
     P^{(c)}_{62} \\P^{(c)}_{71} & P^{(c)}_{72}  \end{array}
  \right) \left( \begin{array}{c}  J_{ab}(s) \\ J_{cd}(s)
     \end{array} \right) \co
\end{align}
where $m_a,M_{\pi^b}$ denote the masses of the final state of the pertinent
channel and $m_c,M_{\pi^d}$ stands for the masses of the intermediate
state that differ from the final state masses. This means for channel $(p0)$
$m_a = m_p, m_c = m_n, M_{\pi^b} = \mpn$ and $M_{\pi^d} = \mpc$.
The elements $P^{(c)}_{ik}$ are functions of the pion momentum $\vq$,
the photon virtuality $k^2$ and the coupling constants of the
Lagrangian. For the channel $(p0)$, one finds
\begin{align}\label{eq:1loopcoef}
P_{11}^{(p0)} &= G_0 C_0(1+\ratio)  + \vq^2 \left(C_0 E^t_{0+,q} - 2 D_0^{(2)}
  G_0\right)+ \frac{k^2}{\alpha_{0,0}^2} C_0 E^t_{0+,k}  \co \nn
  P_{12}^{(p0)} &= C_1 H_0 \left( 1+ \ratio\right) 
  + \qstar{2} \left(C_1 E^{t,(n+)}_{0+,q} -D_1^{(2)} H_0 \right)-\vq^2
  D_1^{(2)} H_0 + \frac{k^2}{\alpha_{0,0}^2} C_1 E^{t,(n+)}_{0+,k}  \co\nn
18 P_{21}^{(p0)} &= \vq^2 \left(D^{(1)}_0-D^{(3)}_0\right)( G_1-G_3) \co\nn
18 P_{22}^{(p0)} &=  \qstar{2} \left(D^{(1)}_1-D^{(3)}_1 \right) (
H_1-H_3) \co \nn
9 P_{31}^{(p0)} &= \vq^2 \left(D^{(1)}_0+2D^{(3)}_0 \right)(G_3-G_1-3G_2) \co\nn
9 P_{32}^{(p0)} &= \qstar{2} \left(D^{(1)}_1+2D^{(3)}_1\right)(H_3-H_1-3H_2)\co \nn
18 P_{41}^{(p0)} &= \vq^2 \left(D^{(1)}_0-D^{(3)}_0\right)(G_1+G_3) \co\nn
18 P_{42}^{(p0)} &= \qstar{2}
\left(D^{(1)}_1-D^{(3)}_1\right)(H_1+H_3) \co \nn
P_{51}^{(p0)} &= G_0 C_0(1+\ratio)  + \vq^2 \left(C_0 S^t_{0+,q}-2 D_0^{(2)}
  G_0\right)+ \frac{k^2}{\alpha_{0,0}^2} C_0 S^t_{0+,k}  \co \nonumber
\end{align}
\begin{align}
P_{52}^{(p0)} &= C_1 H_0(1+\ratio)  + \qstar{2} \left(C_1 S^{t,(n+)}_{0+,q}-D_1^{(2)} H_0 \right)-\vq^2
  D_1^{(2)} H_0 + \frac{k^2}{\alpha_{0,0}^2} C_1 S^{t,(n+)}_{0+,k}  \co\nn
18 P_{61}^{(p0)} &= \vq^2 \left( D_0^{(1)}-D_0^{(3)} \right) (G_1+G_3)  \co \nn  
18 P_{62}^{(p0)} &= \qstar{2} \left(D_1^{(1)}-D_1^{(3)} \right)
(H_1+H_3)  \co\nn
9 P_{71}^{(p0)} &= \vq^2 \left( D_0^{(1)}+ 2D_0^{(3)} \right) (G_1+G_3+3G_{20})  \co \nn  
9 P_{72}^{(p0)} &= \qstar{2} \left(D_1^{(1)}+2D_1^{(3)} \right) (H_1+H_3+3H_{20})  \co
\end{align}
where $E^{t,(c)}_{0+,x}$ denotes the pertinent coefficient of the tree
level result of channel $(c)$, see Eq.~(\ref{eq:tree}), and $\qstar{2}$ is given by
\begin{align}
\qstar{2} &= \frac{\left(s-(m_c+M_{\pi^d})^2\right)\left(s-(m_c-M_{\pi^d})^2\right)}{4s} \co
\end{align}
which is a quantity of order $\epsilon^2$. Up to the order considered
here, the corrections in $k^2$ only contribute to $\Fnr_1$ and
$\Fnr_6$ and they are independent of the scattering angle
$\cos\Theta$. This is the reason why only the coefficients of the
$S$-wave multipoles depend on $k^2$. Eq.~(\ref{eq:1loop}) and
(\ref{eq:1loopcoef}) clearly show the advantage of the nonrelativistic
description: At leading order, the strength of the cusp in the channel $(p0)$ is parameterized in terms of the
coupling constant $C_1$ (i.e. the scattering length $a_{0+}^-$) and the ratio $H_0/G_0$.

{\bf 9.} The two-loop corrections all have the topology shown in
Fig.~\ref{fig:diags} and can therefore be cast into the form
\begin{align} \label{eq:2loop}
E_{0+}^{\mathrm{2 Loop}}(s) &= \big( J_{ab}(s) \,\,\, J_{cd}(s) \big) \left( \begin{array}{cc}  T^{(c)}_{11} & T^{(c)}_{12}\\ T_{12}^{(c)}  & T^{(c)}_{22}  \end{array}\right) \left( \begin{array}{c}  J_{ab}(s) \\ J_{cd}(s)
     \end{array} \right) \co \nonumber \\[2ex] 
S_{0+}^{\mathrm{2 Loop}}(s) &= \big( J_{ab}(s) \,\,\, J_{cd}(s) \big) \left( \begin{array}{cc}  V^{(c)}_{11} & V^{(c)}_{12}\\ V_{12}^{(c)}  & V^{(c)}_{22}  \end{array}\right) \left( \begin{array}{c}  J_{ab}(s) \\ J_{cd}(s)
     \end{array} \right) \fs
\end{align}
The coefficients $T^{(c)}_{ij}$ for the electric multipole in the channel $(p0)$ read
\begin{align}
T_{11}^{(p0)} &= C^2_0 G_0(1+\ratio)  +C^2_0 E^t_{0+,q}  \vq^2  -4 C_0 G_0
D_0^{(2)} \vq^2 +C_0^2 E^t_{0+,k} \frac{k^2}{\alpha_{0,0}^2}   \co\nn
T_{12}^{(p0)} &= \frac{1}{2}\left( C_1^2\,G_0 + C_0 C_1 H_0\right) (1+\ratio) + 
   \frac{1}{2}C_1^2\, E^t_{0+,q}
  \vq^2 - C_1\,H_0\, D_0^{(2)}\vq^2 \nn & -  C_1\,G_0\,D_1^{(2)} \vq^2 -
      \frac{1}{2} C_0\,H_0\,  D_1^{(2)}\vq^2 + 
  \frac{1}{2} C_0\,C_1\,E^{t,(n+)}_{0+,q}\,
     \qstar{2} \nn &- 
  C_1\,G_0\,D_1^{(2)} \, \qstar{2}
  - \frac{1}{2} C_0\,H_0\,
     D_1^{(2)}\, \qstar{2} + \frac{k^2}{2\alpha_{0,0}^2} C_1 \left(C_1
     E^t_{0+,k}  + C_0 E^{t,(n+)}_{0+,k} \right) \co\nonumber
\end{align}
\begin{align}
T_{22}^{(p0)} &= C_1 C_2 H_0(1+\ratio)  - C_2 H_0   D_1^{(2)}\vq^2 +
    C_1 C_2  E^{t,(n+)}_{0+,q} \qstar{2} \nn  &-
    C_2  H_0 D_1^{(2)} \qstar{2} -
    2 C_1 H_0 D_2^{(2)} \qstar{2} + C_1 C_2 E^{t,(n+)}_{0+,k} \frac{k^2}{\alpha_{0,0}^2}  \fs
\end{align}
For the scalar multipole, the coefficients are
\begin{align}
V_{11}^{(p0)} &= \frac{1}{3} C^2_0 G_7 \vq^2-\frac{1}{3} C_0^2 G_8
\vq^2 + \frac{1}{3} C_0^2 G_{21} \vq^2 \co \nn
V_{12}^{(p0)} &= \frac{1}{2}\left( C_1^2\,G_0+ C_0 C_1 H_0 \right) (1+\ratio) + \frac{1}{2}C_1^2\, S^t_{0+,q}
  \vq^2 - C_1\,H_0\, D_0^{(2)}\vq^2 \nn & -  C_1\,G_0\,D_1^{(2)} \vq^2 -
      \frac{1}{2} C_0\,H_0\,  D_1^{(2)}\vq^2 + 
  \frac{1}{2} C_0\,C_1\,S^{t,(n+)}_{0+,q}\,
     \qstar{2} \nn &- 
  C_1\,G_0\,D_1^{(2)} \, \qstar{2}
  - \frac{1}{2} C_0\,H_0\,
     D_1^{(2)}\, \qstar{2} + \frac{k^2}{2\alpha_{0,0}^2} C_1 \left(C_1
     S^t_{0+,k}  + C_0 S^{t,(n+)}_{0+,k} \right) \co\nn
V_{22}^{(p0)} &= C_1 C_2 H_0 (1+\ratio)  - C_2 H_0   D_1^{(2)}\vq^2 +
    C_1 C_2  S^{t,(n+)}_{0+,q} \qstar{2} \nn  &-
    C_2  H_0 D_1^{(2)} \qstar{2} -
    2 C_1 H_0 D_2^{(2)} \qstar{2} + C_1 C_2 S^{t,(n+)}_{0+,k} \frac{k^2}{\alpha_{0,0}^2}  \fs
\end{align}
The two-loop corrections up to and including order $O(a^2
\epsilon^5)$ are independent of the scattering angle and therefore only contribute to the
$S$-wave multipoles $E_{0+}$ and $S_{0+}$. 

{\bf 10.} Given the expressions of the various multipoles, one readily
finds that the difference $E_{1+}-S_{1+}$ is, to
a very high accuracy, free of loop corrections and therefore a
polynomial in $k^2$ and $\vq^2$,
\begin{align}\label{eq:let}
\frac{1}{\vq}\left( E_{1+}-S_{1+} \right)  &= \frac{1}{2} E_{1+}^t \frac{k^2}{\alpha_{0,0}^2} +
\frac{1}{15} (G_{12}-G_{14}) \vq^2 + O(\epsilon^4, a \epsilon^5, a^2
\epsilon^6, a^3 \epsilon^3) \fs
\end{align}
Up to the corrections of higher order, this equation can be
rewritten as a (very) low-energy theorem which relates the derivative of
$E_{1+}$ with respect to $|\vq|$ evaluated at threshold to a
combination of $P$- and $E$-waves,
\begin{align}\label{eq:let2}
k^2\frac{\mathrm{d}}{\mathrm{d}|\vq|} E_{1+}\Big|_{s = s_\mathrm{thr.}}
&= \frac{\alpha_{0,0}^2}{|\vq|}\left(2 E_{1+}-2 S_{1+}+7E_{3+} - 3M_{3+}
+3M_{3-} \right) \fs
\end{align}
From Eq.~(\ref{eq:let}), it is clear that this relation holds for all
values of $\vq$ and $k^2$ which lie in the region of validity of the
effective theory, in particular for $\vq = 0$ since each
of the multipoles on the right hand of
Eq.~(\ref{eq:let2}) is proportional to $|\vq|$.

{\bf 11.} The result for the other channels are obtained from the tree
level result of channel $(p0)$ and the coefficients $P_{ij}^{(p0)}$ and $T_{ij}^{(p0)}$
by the replacements
\begin{align}
(n+) &: \{G_i,H_i,C_0,C_2 \} \to
  \{H_i,G_i,C_2,C_0 \}\co \nn
(n0) &: \{G_i,H_i,C_0,C_1,C_2 \}   \to
  \{L_i,K_i,C_3,C_4,C_5 \} \co \nn
(p-) &: \{G_i,H_i,C_0,,C_1,C_2 \}   \to
  \{K_i,L_i,C_5,C_4,C_3 \} \fs
\end{align}
The replacement indicated for the $C_x$ has to be done also for the
corresponding $D^{(i)}_x$. 

{\bf 12.} In this letter, the electroproduction reaction of pions on the 
  nucleon is studied using a nonrelativistic framework. The electric,
  scalar and magnetic multipoles $E_{l+}, S_{l+}$ for $l = 0,1$ and $M_{1\pm}$ are
  calculated in a systematic double expansion in the final state pion-
  and nucleon momenta normalized with the pion mass (counted as a small quantity of order
  $\epsilon$) and the photon virtuality $k^2$ as well as
  the threshold parameters of $\pi N$ scattering. Explicit representations for the multipoles up to and
  including $\epsilon^3$ at tree level, $\epsilon^4$ at one loop and
  $\epsilon^5$ at two loops are
  provided.  The effective theory expansion shows a good convergence
  behavior in the low-energy region,
  at least up to a pion momentum of $|\vq| \simeq 70 \MeV$ and for photon virtualities which satisfy $|k^2| \ll 4\mpc^2$. It
  accurately describes the cusp structure and allows one to
  determine the pion-nucleon threshold parameters from experimental
  data. As an application, a new low-energy theorem relating the slope
  of $E_{1+}$ at threshold to a combination of $P$- and $E$-wave
  multipoles is discussed. \newline
  A numerical analysis of all available experimental data of pion
  electro- {\it and} photoproduction in order to determine the pion-nucleon threshold
  parameters will be presented in a
  subsequent publication. 

{\it Acknowledgments.} I would like to thank J.~Gasser for
comments on the manu-script. This work was supported in part by the
Department of Energy under Grand DE-FG03-97ER40546.

\renewcommand{\thefigure}{\thesection.\arabic{figure}}
\renewcommand{\thetable}{\thesection.\arabic{table}}
\renewcommand{\theequation}{\thesection.\arabic{equation}}

\appendix

\setcounter{equation}{0}
\setcounter{figure}{0}
\setcounter{table}{0}

\section{Matching relations}\label{app:matching}

The matching relations of the nonrelativistic couplings to the
threshold parameters defined in Eq.~(\ref{eq:thresholdpar}) read
\begin{align}
2G_4 &= 15 (\bar{E}_{2+}+2 \bar{M}_{2+}) \co \nn
2G_5 &= 2\zeta G_0- 2\bar{E}_{2-}- 
2\bar{E}_{0+,2,0}+3 (\bar{E}_{2+}-2
  \bar{M}_{2-}+6\bar{M}_{2+}) \co \nn
G_6 &= -3 (3\bar{M}_{2+}+2\bar{M}_{2-}) \co \nn
G_7 &= 15 (\bar{E}_{2+}-\bar{M}_{2+}) \co  \nn
G_8 &= 3 (\bar{E}_{2-} -
  \bar{M}_{2+}+\bar{M}_{2-}+\bar{E}_{2+})\co \nonumber
\end{align}
\begin{align}
2 G_9 &= 4G_1 \zeta - 6\bar{E}_{3-} -
6\bar{E}_{1+,2,0}+15\bar{E}_{3+}-24\bar{M}_{3-}-6 \bar{M}_{1+,2,0} +45 \bar{M}_{3+}\co \nn
2 G_{10} &= 35 (\bar{E}_{3+}+3\bar{M}_{3+})\co \nn
2G_{11} &= 2 G_2 \zeta +  2\bar{M}_{1-,2}-
9\bar{M}_{3-}+ 4 \bar{M}_{1+,2} -12\bar{M}_{3+} \co \nn
2G_{12} &= -15 (3\bar{M}_{3-}+4\bar{M}_{3+}) \co  \nn
2G_{13} &= 4 G_3 \zeta -6 \bar{E}_{3-}-6
\bar{E}_{1+,2,0}+15\bar{E}_{3+}-6 \bar{M}_{3-} + 6 \bar{M}_{1+,2,0}- 15 \bar{M}_{3+}\co \nn
2G_{14} &= 105 (\bar{E}_{3+}-\bar{M}_{3+}) \co  \nn
G_{15} &= 15
  (\bar{M}_{3-}+\bar{E}_{3-}+\bar{E}_{3+}-\bar{M}_{3+})\co\nn
2 G_{16} &= 2\alpha_{0,0}^2 \bar{E}_{0+,0,2} - G_0 -2G_0 \alpha_{0,0}
\alpha_{0,1} \co \nn
2 G_{17} &= 6 \alpha_{0,0}^2 \left(\bar{E}_{1+,0,2}+\bar{M}_{1+,0,2}
\right)-G_1 - 4 G_1 \alpha_{0,0} \alpha_{0,1} \co \nn
2 G_{18} &= 6 \alpha_{0,0}^2 \left(\bar{E}_{1+,0,2}- \bar{M}_{1+,0,2}
\right) - G_3 + 4 G_3 \alpha_{0,0} \alpha_{0,1} \co \nn
G_{19} &= -\alpha_{0,0}^2 \left(\bar{M}_{1-,0,2}+ 2\bar{M}_{1+,0,2}
\right)- G_2 \alpha_{0,0} \alpha_{0,1} \co \nn
G_{20} &= \bar{S}_{1-}-2\bar{S}_{1+} \co \nn
G_{21} &= 6 \bar{S}_{2-}- 9 \bar{S}_{2+}+G_8\co \nn
2 G_{22} &= 9 \bar{S}_{3-} -12 \bar{S}_{3+} -2 \bar{S}_{1-,2,0} +4
\bar{S}_{1+,2,0}+ 4 \zeta G_{20} \co \nn
G_{23} &= \alpha_{0,0}^2 \left(\bar{S}_{1-,0,2}-2\bar{S}_{1+,0,2}
\right) -2 G_{20} \alpha_{0,0} \alpha_{0,1} \co
\end{align}
with $\zeta = \frac{\alpha_{1,0}}{\kt}$.

\end{document}